# The Steward Observatory LEO Satellite Photometric Survey

Harrison Krantz , Eric C. Pearce, and Adam Block
University of Arizona Steward Observatory, USA
*Received 2023 January 30; accepted 2023 August 25; published 2023 September 21*

## Abstract

The Steward Observatory LEO Satellite Photometric Survey is a comprehensive observational survey to characterize the apparent brightness of the Starlink and OneWeb low Earth orbit satellites and evaluate the potential impact on astronomy. We report the results of over 16,000 independent measurements of nearly 2800 individual satellites. In addition to photometry, we also measured the astrometric position of each satellite and evaluated the accuracy of predicting satellite position with the available two-line element sets. The apparent brightness of a satellite seen in the sky is not constant and depends on the Sun-satellite-observer geometry. To capture this, we designed the survey to create an all-geometries set of measurements to fully characterize the brightness of each population of satellites as seen in the sky. We visualize the data with sky-plots that show the correlation of apparent brightness with on-sky position and relative Sun-satellite-observer geometry. The sky-plots show where in the sky the satellites are brightest. In addition to visual magnitudes, we also present two new metrics: the expected photon flux and the effective albedo. The expected photon flux metric assesses the potential impact on astronomy sensors by predicting the flux for a satellite trail in an image from a theoretical 1 m class telescope and sensor. The effective albedo metric assesses where a satellite is more reflective than baseline, which ties to the physical structure of the satellite and indicates the potential for brightness-reducing design changes. We intend to use this methodology and resulting data to inform the astronomy community about satellite brightness. Observing programs use a variety of telescopes and instruments and look at different parts of the sky. With the expected photon flux metric and a complete all-sky characterization of satellite brightness, observers can evaluate the potential impacts to their projects and possibly avoid the worst effects.

*Unified Astronomy Thesaurus concepts:* Artificial satellites (68); Photometry (1234); Astrometry (80); Astronomical site protection (94); Ground-based astronomy (686); Astronomical instrumentation (799)

## 1. Introduction

From 2017 to 2022, the number of intact space objects (satellites and rocket bodies) in Earth orbit more than doubled, with most of the growth happening in just three years (see Figure 1). The increasing number of space objects is concentrated in low Earth orbit (LEO) and primarily driven by deployment of new large satellite constellations, each consisting of hundreds or thousands of individual satellites. These satellite constellations provide a variety of services including navigation, communications, and remote sensing. Particularly notable are the communications constellations Starlink and OneWeb, which as of the end of 2022 are the two largest constellations and together outnumber all other active satellites in orbit. However, these are only two of many companies that plan to create their own communications satellite constellations. In total the future may include as many as one hundred thousand satellites in LEO (Williams et al. 2021). This is an exciting future for space development but carries with it multiple new challenges including space traffic management (Hiles et al. 2021), conjunction and collision risks, environmental impact from rocket launches and satellite atmospheric disposal (Boley & Byers 2021), and incidental effects to astronomy (JASON Program Office 2020; Walker et al. 2020b). The focus for this paper is the incidental effects to astronomy.

When an orbiting satellite is illuminated by sunlight and over an area of Earth where it is night, it may be visible in the night sky and appear as a star-like point source moving across the sky. Satellites in high orbits, such as geosynchronous, are nearly continuously illuminated and visible all night. Satellites in lower orbits, such as the Starlink and OneWeb satellites, periodically pass through the Earth's shadow as shown in Figure 2 and are not visible to optical telescopes while eclipsed. Most LEO satellites are only illuminated and visible during twilight and the early hours of night, with the exact duration of visibility dependent on the satellite's orbit, latitude of the observer, and season. Satellites in a circular 500 km orbit, like Starlink, may be visible up to four hours after sunset from a mid-latitude location and all night from a high-latitude location during summer. Lawler et al. (2022) presents a comprehensive look at the visibility of large satellite constellations at various latitudes.

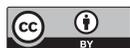

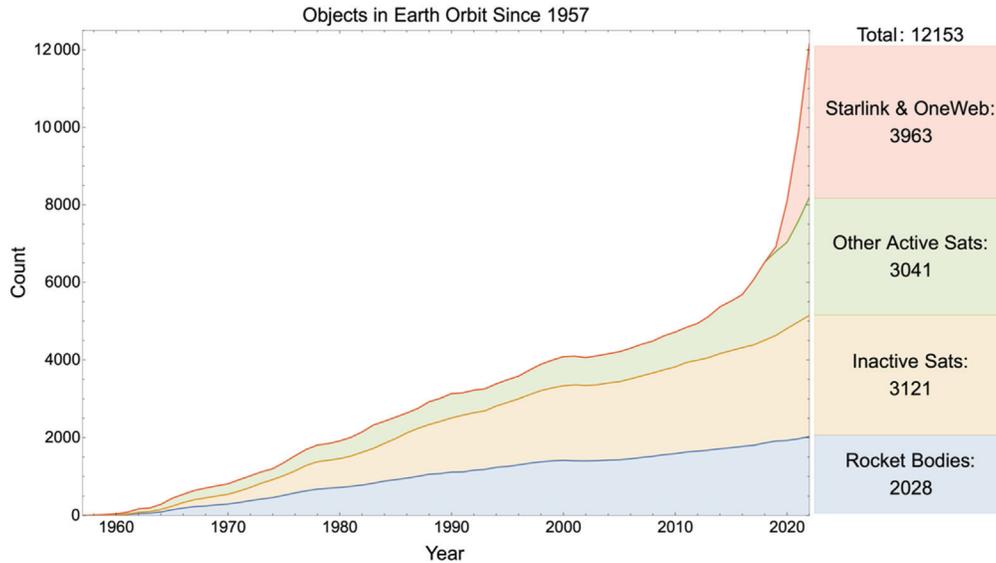

**Figure 1.** The number of intact space objects (satellites and rocket bodies) in orbit more than doubled from 2017 to 2022. This growth is primarily driven by new large satellite constellations including Starlink and OneWeb which outnumber all other active satellites. Data compiled from the public US Satellite Catalog available via Space-Track.org and from GCAT (J. McDowell, planet4589.org/space/gcat).

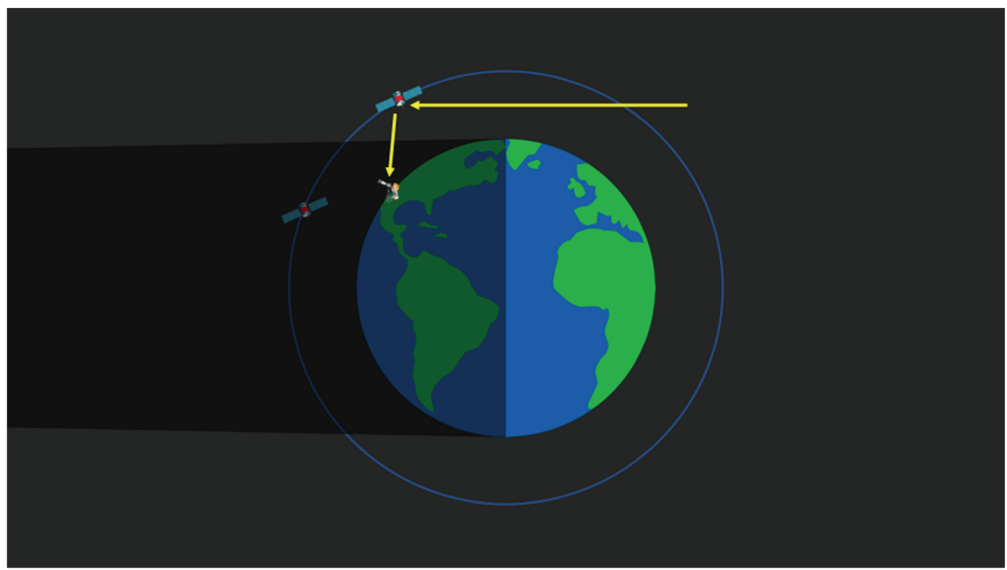

**Figure 2.** A satellite illuminated by sunlight may be visible to observers on the ground where it is night. Conversely, a satellite that is eclipsed by Earth's shadow is not illuminated by sunlight and not visible to optical telescopes.

A bright satellite may be visible to the unaided eye, but many more are visible to telescopes and cameras that have higher sensitivity and can see fainter objects. For astronomers these satellites can be a nuisance to stargazing, astrophotography, and scientific research. As a point of reference, a 1 m sphere with an albedo of 20% at 500 km will have a visual magnitude of approximately 6.7 at zenith during twilight (Hejduk 2011). The new LEO satellites are as bright at 4th magnitude, and in some cases even brighter. This is up to a million times brighter than many typical faint astronomical objects.

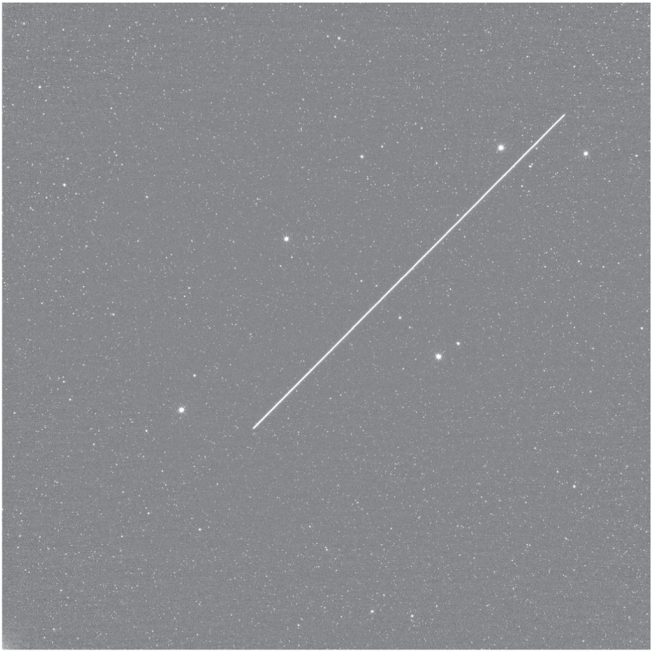

**Figure 3.** A 3 s exposure image from the Steward Observatory SSA astrograph with a 4.2° × 4.2° field of view showing a trail from STARLINK-1284 which measures as 1.80 mag.

In images, satellites appear as lines or trails like the example shown in Figure 3, potentially obscuring a target of interest or otherwise marring the image. Depending on the use-case, a satellite trail in an image may not be a hindrance (e.g., if the target of interest is not obscured) or can otherwise be corrected with software. One common software method is to stack multiple images and compute the median or mean values improving the signal-to-noise ratio and removing outlier values including those created by satellite trails. In some cases, the images with satellite trails are discarded entirely because the satellite trail irrecoverably affected the data or recovering the data is otherwise too difficult. This is common when the image with a satellite trail is only one of many images and discarding one image is not a significant detriment. Very sensitive instrumentation may exhibit secondary effects from bright satellite trails, such as ghosting or the electronic crosstalk seen with the Vera C. Rubin Observatory LSST Camera (Tyson et al. 2020).

Illuminated and visible satellites in the night sky are not new and astronomers have worked around them for decades. Until recently, seeing a satellite was a familiar but uncommon event. Now with the rapidly increasing number of satellites in orbit, there are more satellites visible in the night sky, and more images incidentally recording satellite trails. Additionally, the new communications constellation satellites are brighter than the previous era of satellites. The brighter appearance is a consequence of the low orbits utilized by the new satellite constellations. Previously, the majority of illuminated satellites were in higher orbits, such as geostationary, and are much fainter and less impactful than the new communications constellation satellites in LEO.

Satellite brightness is a complicated and underappreciated phenomenon. Satellite brightness is not constant and depends on many variables relating to the illumination and reflection geometry. The apparent brightness for an individual satellite is the combination of reflected light from all components of the satellite. Each component produces its own reflection with both diffuse and specular properties. Variations in satellite orientation or the angle of oncoming sunlight change the reflected light from each component and cause the apparent brightness as seen by an observer to change. The first-generation Iridium satellites famously flared in brightness when their polished antennae panels specularly reflected sunlight at specific angles.

Most important is the Sun-satellite-observer geometry, which comprises the angle of oncoming sunlight and reflecting angle to the observer. For an observer on the ground the Sun-satellite-observer geometry includes a large range of possible angles as the satellite can be anywhere in the hemispherical sky. The apparent brightness of a satellite will be different depending on where in the sky it is seen and the corresponding Sun-satellite-observer geometry.

With all other factors equal, a satellite will appear fainter with increasing distance from the observer. However, when recorded in an image, the brightness of the satellite trail depends on both the instantaneous brightness of the satellite and its apparent on-sky motion which depends on its orbit and distance from the observer. A far-away slow-moving faint satellite may appear just as bright, per pixel, as a closer faster-moving brighter satellite. Section 4.2 includes more discussion on this and introduces a new metric to evaluate satellite brightness as seen by an imaging sensor.

Impacts on astronomy are not created by singular satellites, or the brightness of individual satellites, but the combined effects of the entire ensemble of visible satellites. Already, communications constellation satellites number in the thousands and are expected to increase to tens of thousands in the future. The appearance of numerous satellites in the night sky will not make astronomy activities impossible but will make them more difficult. More frequent appearance of satellite trails in images will lead to discarding more images or require more effort to recover them with software. Scientific research projects will require more observing time, more processing time, and more funding to complete.

Not all astronomy will be affected equally. Wide-field survey instruments are more likely to incidentally see satellites with their large field of view (FOV) whereas narrow-field instruments are less likely to encounter them. Larger telescopes and more sensitive instruments can see fainter objects, and therefore also see fainter and more satellites. Projects that prioritize twilight observing time, such as searches for near-Earth objects, observe when the most satellites are illuminated and visible whereas projects that observe around midnight avoid the majority of visible satellites. The exact scope of effects and overall impact depend on the observational project and specific instrumentation. Radio astronomy is impacted too, however in a different manner than optical astronomy which the focus of this paper. While satellites appear as finite point sources in the FOV of optical observatories (and only during times when the satellites are illuminated and visible), radio observatories are susceptible to interference from satellites anywhere in the sky and during all times of night and day.

To discuss concerns regarding the impacts of satellites on astronomy, the community convened several meetings in 2020 and 2021, including the SATCON 1 and 2 (Walker et al. 2020b; Hall et al. 2021) and Dark and Quiet Skies 1 and 2 (Walker et al. 2020a, 2021) workshops. In 2022, the International Astronomical Union formed the Centre for Protection of the Dark and Quiet Sky from Satellite Constellation Interference (IAU CPS) (IAU 2022). Through the meetings, and now the IAU CPS, astronomers, satellite operators, and additional communities are working to understand the impacts and create solutions to mitigate them.

Key to understanding the impact on astronomy is determining how bright the satellites are in the night sky and their frequency of appearance in the field of regard relevant to the particular observatory and observing program. With knowledge of satellite brightness and distribution, astronomers can determine if satellites will be visible to their instruments, how they will appear in their data, and assess the level of impact on their project. With a baseline measurement of satellite brightness, the community can evaluate the efficacy of current mitigation efforts and set goals for future efforts.

Members of the astronomy community have published articles reporting the brightness of communications constellation satellites (Horiuchi et al. 2020; Mallama 2020, 2021; Tyson et al. 2020; Tregloan-Reed et al. 2021; Boley et al. 2022; Halferty et al. 2022; Mróz et al. 2022). These reports provided valuable information to the community asking how bright these satellites are. However, these reports do not fully describe the satellite brightness due to limited measurements, either in number or geometric diversity, or do not describe the brightness correlations with all variables as seen in the sky, across the whole sky, in a way that is relevant to astronomers. Many modified their measurements to a standard range for comparison and in doing so create numbers which do not describe the brightness as an observer will see the satellite. Assessing the impact on astronomy requires a complete view of satellite brightness as seen in the sky.

To measure and characterize the brightness of communications constellation satellites across the entire range of possible geometries we created the Steward Observatory LEO Satellite Photometric Survey. This automated survey observes satellites to measure their apparent brightness and astrometric position. With an all-sky, all-geometries set of measurements we fully characterize the range of brightness for each population of satellites as seen in the sky. We intend to inform the astronomy community how bright they can expect communications constellation satellites to be in different parts of the sky. With this information, observers can evaluate the potential impact on their projects and possibly avoid the brightest satellites and worst effects.

## 2. Methods

Unlike typical targets of astronomical observing, satellites move across the sky with significant angular rates relative to the celestial sphere and require adaptation of conventional techniques to observe. In particular, LEO satellites move with varying angular on-sky velocities up to 1°–2° per second and are only above the horizon for 10–20 minutes at a time. While observing a moving target is simple in description, in practice it is much more challenging. There are two methods of observing a fast-moving satellite: tracking and snapshot (also referred to as Wait & Catch).

Tracking fast-moving LEO satellites is difficult to do well and requires a fast and accurate mount. Tracking the satellite and recording consecutive images produces a lightcurve showing changes in brightness as the satellite passes overhead and goes through a range of Sun-satellite-observer geometries. Photometry of the satellite is simple because it appears star-like in appearance, i.e., a single confined point source. However, the background stars are severely streaked making conventional photometric and astrometric techniques difficult. Photometric calibration is possible using relative or all-sky photometric techniques, but these measurements are time consuming for survey work.

With the snapshot method, instead of tracking the satellite, the telescope sidereally tracks a point in the sky where the satellite will be. Then at the precise moment when the satellite is crossing the FOV the camera records an image of the satellite. Reliably executing observations in this way requires planning and precise execution timing of both the telescope and camera system. This method produces an image with static point source background stars and a satellite trail. If the entire satellite trail is wholly contained within the image frame like in Figure 3, then photometry is simply the sum of the flux in the trail and the astrometric position is the centroid of the trail, using background stars for reference. It is possible to perform photometry with a truncated satellite trail (Tregloan-Reed et al. 2021), though this requires estimating the satellite's apparent velocity from its orbit.

For our survey we utilized the snapshot method. The snapshot method is easier to automate with our equipment, produces data that is easier to process, and requires less time per individual satellite allowing the survey to observe more satellites and produce a more diverse data set. We do track

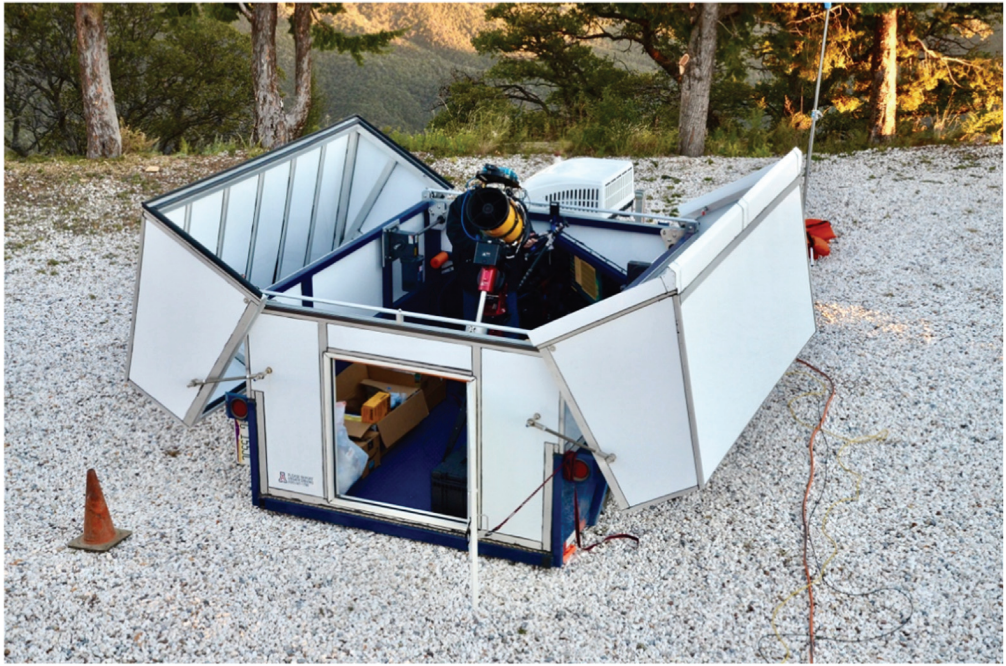

**Figure 4.** The Steward Observatory SSA astrograph and the unique portable trailer-mounted enclosure. The astrograph has a 180 mm aperture and 4.2° × 4.2° FOV.

specific satellites on a limited basis, but these measurements are not discussed in this paper.

For either method, predicting where a satellite will be requires information about its orbit. This information is most readily available as a Two-Line Element Set (TLE). This is a legacy format consisting of two lines of 69 characters that include the satellite ID number, epoch, and orbital elements (Space-Track.org 2022). When input into a simplified perturbation model (e.g., SGP4) the position of the satellite is computed for a specific time (Vallado et al. 2006). Due to variations in the space environment (e.g., drag) and general uncertainty, a specific TLE is only accurate for a limited time near the epoch of its creation and loses accuracy as the time since the epoch increases. New TLEs are regularly created from observational monitoring and satellite telemetry data. The United States Space Force publishes a public catalog of TLEs through Space-Track.org. Separately, Celestrak.org publishes Supplemental TLEs for some satellites, including Starlink and OneWeb satellites, which are generated from first-party telemetry data provided by satellite operators (Celestrak 2022).

### 2.1. Instrumentation

The Steward Observatory SSA astrograph, shown in Figure 4, is a unique system specifically created to observe Earth satellites and space debris. The telescope is a 180 mm Takahashi Epsilon hyperbolic Newtonian astrograph, which provides a 4.2° × 4.2° FOV at 4.95″ pixel$^{-1}$ on a 3056 × 3056 CCD Apogee Alta F9000 imager with a 7-color filter wheel. The survey utilized an Astrodon Johnson-Cousins $V$ filter for all observations. The telescope rides on a Paramount MYT German equatorial mount (Pearce et al. 2018).

The system is fully robotic, allowing for remote operation and automated observing. ACP Observatory Control software executes automated observing operations (Denny 2022). The system resides in a unique portable trailer-mounted enclosure (Krantz 2022) with weather monitoring and sky brightness sensors. The mobile enclosure enables relocation for different observing programs with minimal necessary site infrastructure. For the majority of the survey, we operated on the grounds of the Mt. Lemmon Sky Center, at the summit of Mt. Lemmon near Tucson AZ. We temporarily relocated the system to the Biosphere 2, also near Tucson AZ, from 2020 August to mid-October due to the Bighorn wildfire which forced closure and evacuations from Mt. Lemmon.

The robotic operation and large FOV of the astrograph make it particularly capable of observing fast-moving satellites. The reliable timing of robotic operation enables observation of many satellites in quick succession. The large FOV allows for a large tolerance of error in observation timing and uncertainty in the satellite position. Furthermore, the large FOV captures the entire satellite trail in the image frame with a reasonable exposure time for reference background stars, simplifying the photometry and reducing measurement uncertainty. With a smaller FOV, the satellite trail may be truncated by the image frame and the apparent brightness is not simply determined from the sum of flux in the trail.

We utilize cloud computing resources to manage the survey. A MySQL database holds a record of satellites, TLEs, and observational data. Custom software regularly updates the database with new satellites and TLEs. We archive both the standard issue TLEs and Celestrak's Supplemental TLEs.

### 2.2. Observation Scheduling

Satellites are constantly orbiting Earth and regularly passing overhead. From our location near Tucson, AZ, our target LEO satellites are only visible in the night sky for a finite time twice each day during twilight hours when they are illuminated by sunlight. For scheduling, this is defined as two observing time windows: evening and morning. We created a scheduler software program to determine when satellites will be visible, select a subset to observe, and create an observing script to execute. This software program is written in Python and utilizes the Skyfield library (Rhodes 2019).

The scheduler program runs automatically every evening before sunset. The program begins by querying the database for the latest TLE for each satellite. It then computes for every satellite the position and visibility in the observing time window. The time window for evening observing begins 45 minutes after sunset and ends when no more satellites are visible, and vice versa for morning observing. The duration of this time window varies with the season from only two hours in the winter to as long as four hours in the summer. The program selects satellites that are above the horizon limit (20°) and illuminated by sunlight.

Due to the large number of visible satellites, it is not possible to observe all in one observing session, so the program selects a subset to observe based on a priority weighting scheme. First the program randomly sets a time to observe each visible satellite during its visibility window. It then constructs a graph (the abstract data structure) of all the visible satellites represented as nodes and weights equal to the priority level. Edges represent chronology where two satellite nodes are only connected if the time duration between them is longer than the required overhead time between observations (set to 90 s). This time allows for the telescope to slew and perform other operations between each observation. Additionally, the program adds two dummy nodes to the graph, one for the beginning of the observation time window and one for the end. The scheduler program then uses an iterative optimization algorithm to compute the path through the graph that accumulates the highest sum of weights by visiting various nodes. With this method a higher priority satellite will only be observed if its priority value is higher than the sum of priority values of other satellites that could be observed instead.

After determining the optimal schedule, the scheduler program writes the selected satellites and ephemeris information to an output file and writes an "ACP Plan" script file which will later drive the robotic telescope and camera.

By randomizing the selected time to observe each satellite the survey effectively randomly samples different positions in the sky in correlation with the visibility of satellites in the sky. The survey observes more satellites low in the sky because there are more satellites visible low in the sky. Over time the survey accumulates enough measurements to cover the entire hemispherical sky with a data set that represents the distribution of satellites in the sky.

### 2.3. Observation Execution

Observation execution is controlled with ACP Observatory Control software. This program reads an "ACP Plan" script and runs the telescope and camera systems executing observations in sequence. We created an additional supervisor software program to initiate ACP and start running individual plans at specific times rather than combining all observations for the survey and other projects into one single plan.

ACP is designed for conventional astronomy observations and like many astronomy scheduler and queue programs does not accommodate the specific needs of observing satellites well. In particular is the lack of capability to schedule an observation at a precise time. ACP schedules observations with the "Wait until" command but this command delays all actions of the telescope (i.e., slewing, filter change etc.) not just image acquisition. To reliably observe satellites, we need to trigger the camera at precise times without hindrance of the variable slew time or other overhead. We created a workaround solution to write an "ACP Plan" that will trigger the camera at the specific time. For each individual satellite observation, the script includes a series of actions:

1. Wait until 90 s before the specified observation time with a "Wait until" command.
2. Slew to the specified R.A. and decl. coordinates and record an image (one combined image acquisition command in ACP).
3. Wait until the specified observation time with a "Wait until" command. During this time the telescope continues to track at a sidereal rate.
4. Record an image at the specified R.A. and decl. coordinates. Since the telescope is already at the specified position the only overhead time necessary is that for ACP to check telescope status and command the camera to record an image. The schedule calls for image acquisition 9 s prior to the observation time to account for overhead time.

With this procedure the system records two images for every satellite observation: one clean image before the satellite's arrival and a second image with the satellite trail. Although initially a consequence of actions necessary to make ACP work for our purpose, the clean image is used during image processing to subtract background stars and improve the resulting photometry.

The images are 3 s exposures. We determined this to be a good balance between being long enough to capture background stars for photometric reference and short enough to reliably capture the entire satellite trail within the image frame.

### 2.4. Image Processing

We created our own software pipeline to process the images and accommodate the unique challenges to perform photometry with the elongated satellite trails. Most of the software is written in Python and makes use of Astropy libraries (Astropy Collaboration et al. 2022). We also utilize additional software programs and tools for certain procedures.

#### 2.4.1. Image Reduction

We created a Python program to reduce the raw images and produce calibrated image files. The program follows standard procedures for subtracting the bias frame and dividing by the flat frame. We do not perform dark subtraction due to the short exposure time during which no measurable dark current accumulates. In addition to conventional reduction procedures, the program performs a global background subtraction to remove any background light and ease later procedures in the pipeline. We utilize the Astropy Photutils Background2D function to estimate and map the background flux, which is subsequently subtracted from the image.

We acquire new calibration bias and flat frame data every several weeks. We utilize CCDStack2 to generate master bias and master flat frames, combining multiple raw images with a sigma reject mean algorithm (CCDWare 2022). When reducing the science images, the software program automatically selects the most timely calibration master frames.

#### 2.4.2. Plate Solving

We created a Python program to plate solve each image and add World Coordinate System (WCS) data. This program utilizes Astropy Astroquery to query Astrometry.net for plate solutions (Lang et al. 2010). To minimize network usage the program only submits the pixel coordinates of the 50 brightest stars in the image which are determined using Astropy Photutils Source Detection functions. To minimize computation time, the program provides additional metadata information to Astrometry.net including the pixel scale and approximate R.A. and decl. of the image. The programs adds the WCS solution returned by Astrometry.net to the science image file.

#### 2.4.3. Zero-point Calculation

We use a modified version of The Photometry Pipeline (Mommert 2017) to determine the photometric zero-point for each image. The Photometry Pipeline itself is an entire processing pipeline created to calculate photometry for asteroids and minor bodies. In this use case we utilize The Photometry Pipeline for its automated background star measurements, star catalog queries, and zero-point determinations.

#### 2.4.4. Target Selection

To measure the flux of the satellite trail we first need to detect all the pixels that are part of the satellite trail. We tested multiple algorithms and while many successfully detected the satellite trail in most cases, all required manual review to avoid erroneous or missed detections. The algorithms easily detected bright, well-defined satellite trails but faint trails and edge cases yielded many erroneous and failed detections. The algorithms rarely detected faint trails barely above the noise level or only detected them in parts. Ignoring these faint trails would produce an unwanted bias to only include bright satellites in the final data set. During high winds, vibration in the telescope during image exposure results in a squiggly but otherwise measurable satellite trail. Measuring these examples excludes using straight line detection algorithms. Occasionally multiple satellites trails appear in one image, sometimes overlapping, requiring careful selection of only the targeted satellite.

The satellite trail detection challenges are likely solvable with an appropriate set of algorithms and software development. However, we decided to prioritize data processing over software development and rely on human-based target selection. We created a program that automatically detects all sources in each image, which is then reviewed by a human who manually selects the satellite trail in whole. The program shows each image to the user with the detected sources displayed with different colors. The user selects the satellite trail by clicking on it and changing its color to mark it as the target. In many cases, the program automatically detects the satellite tail as a single source, though in some cases the user must click on multiple adjacent sources to select the entire satellite trail. In cases where the satellite trail is very faint and not automatically detected, the user manually adjusts the detection threshold to recover and detect the trail. If there is no satellite trail, or there are other issues with the image such as significant cloud obscuration, the user flags the image for rejection. After completion, the program saves for each image an image mask, which signifies the pixels selected as the satellite trail.

#### 2.4.5. Photometry

We created a Python program to calculate the photometry for the satellite trails. This program loads the target image, the target image mask, and the corresponding clean image, which includes the same field of background stars. The program applies the target image mask to both the target image and the clean image. The program sums the flux in the unmasked pixels in each image and subtracts the clean image sum from the target image sum to produce the final flux sum for the satellite trail. In many cases the clean image sum is near zero, though in cases where the satellite trail overlapped a background star the clean image sum is non-zero and subtracting it from the target image sum removes the additional flux provided by the background star.

The program calculates the instrumental magnitudes and corresponding statistical uncertainties from the summed flux and writes these values to an output file, which is then uploaded to the database. The final calibrated magnitudes are later calculated in the database via the sum of the instrumental magnitude and zero-point determined for the image.

### 2.4.6. Astrometry

The same program that calculates the photometry also determines the R.A. and decl. coordinates for the centroid of the satellite trail from the WCS solution. This is the real position of the satellite at the mid-point time of the image exposure. A separate program also calculates the expected satellite position using the most timely TLE and compares the coordinates to evaluate TLE accuracy.

## 3. Targets

The Steward Observatory LEO Satellite Photometric Survey observed the Starlink and OneWeb satellite constellations. To best characterize these satellites, we differentiate each model of a satellite and consider each group as its own statistical population.

### 3.1. SpaceX Starlink

There are multiple different models of first generation Starlink satellites all with the same basic design but with various modifications and material differences (SpaceX 2020a, 2020b, 2022). The first generation Starlink satellites utilize a unique L-shaped design, shown in Figure 5. SpaceX created this design to fold flat and compactly stack up to 60 satellites on one rocket for launch. The satellite bus is a flat rectangle approximately 1 m × 3 m in size and utilizes reflective AgFEP (silvered fluorinated ethylene propylene) on the exterior. The bus includes four flat phased array antennae built into the nadir face and two rotatable parabolic dish antennae, one on each end of the bus. A single solar array approximately 10 m long extends from the long edge of the bus forming an overall L-shape. The angle between the bus and solar array is adjustable and changes with operating mode. During the orbit raising phase the satellites operate in a low-drag "open-book" configuration where the bus and solar array are unfolded to form a single flat plane. When at operational orbit the satellites reconfigure to the "shark-fin" configuration where the bus and solar array form an L-shape (SpaceX 2020a, 2020b).

Starlink satellites operate in multiple 540–570 km circular orbital shells. The satellites initially launch into a lower orbit

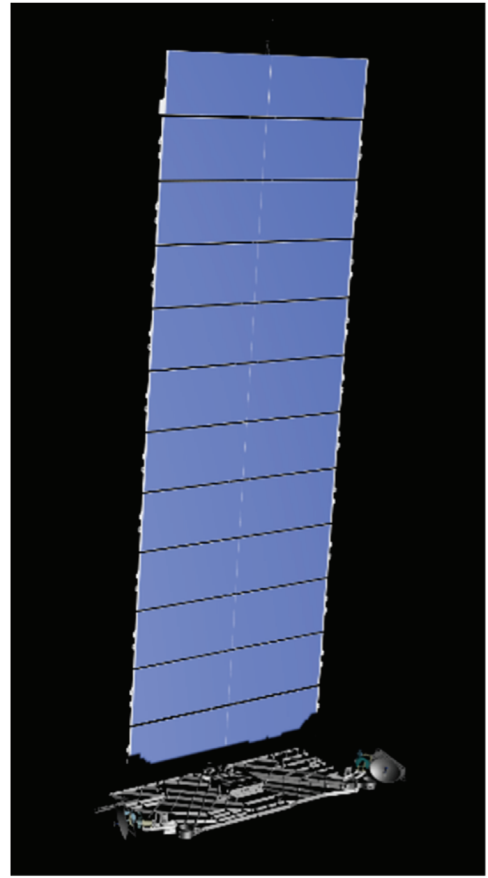

**Figure 5.** The Starlink satellites utilize a unique L-shaped design with a flat rectangular bus and single solar array. Reproduced with permission from SpaceX 2020a. © SpaceX. Copyright 2023.

around 300 km and then individually use on-board thrusters to raise to the final operational orbit.

As of the end of 2022, there are 3460 Starlink satellites in orbit with an expected total of 12,000 satellites to complete the initial constellation. An extended constellation may include as many as 42,000 satellites.

#### 3.1.1. Starlink v0.9

The Starlink v0.9 satellites are prototype versions launched in 2019 May (Krebs 2022a). These prototype satellites have fewer components and different material finishes than later versions; notably missing are the two parabolic dish antennae which are included on later models. As of the end of 2022, all Starlink v0.9 satellites have deorbited.

#### 3.1.2. Starlink v1.0

The Starlink v1.0 satellites are the first production version launched from 2019 November to 2020 June (Krebs 2022b). These satellites utilize diffuse white materials on the nadir-facing phased array antennae (shown in Figure 6), the parabolic dish antennae, and the solar array backing (SpaceX 2022).

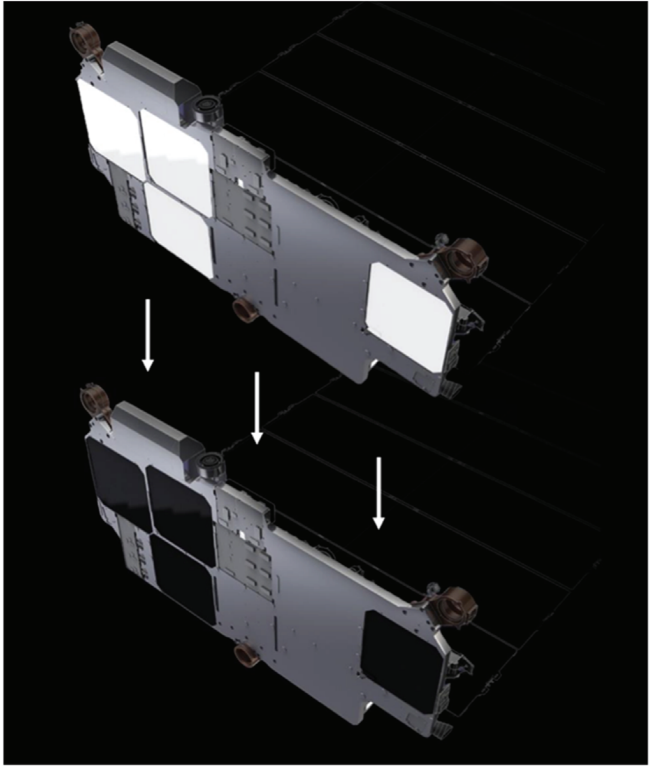

**Figure 6.** Upper: the Starlink v1.0 satellites utilize a diffuse white materal on the nadir-facing phased array antennae. Lower: The Starlink DarkSat prototype utilizes dark materials on the nadir-facing phased array antennae. Reproduced with permission from SpaceX 2020a. © SpaceX. Copyright 2023.

### 3.1.3. Starlink v1.0 DarkSat

Starlink DarkSat is a single prototype satellite launched in 2020 January. This prototype satellite is a modified version of the Starlink v1.0 satellite and utilizes dark material on the nadir-facing phased array antennae (shown in Figure 6) and for covers on the parabolic dish antennae (SpaceX 2020a). Otherwise, we assume this satellite is identical to the other v1.0 satellites.

### 3.1.4. Starlink v1.0 VisorSat

The Starlink v1.0 VisorSat satellites are a modified version of the Starlink v1.0 satellites. The first example is a single prototype satellite dubbed "VisorSat" launched in 2020 May. After success with the prototype, all production satellites launched from 2020 August to 2021 May utilize this design (Krebs 2022b). These satellites include shade visors which hang below the satellite from the edge of the satellite bus (shown in Figure 7). These visors, made from a dark material, are specifically shaped to block sunlight from reaching the nadir-facing phased array antennae. Additionally, the parabolic dish antennae include covers made from dark material like on the Starlink DarkSat prototype. Diffuse white materials are utilized on the nadir-facing antennae and solar array backing (SpaceX 2022).

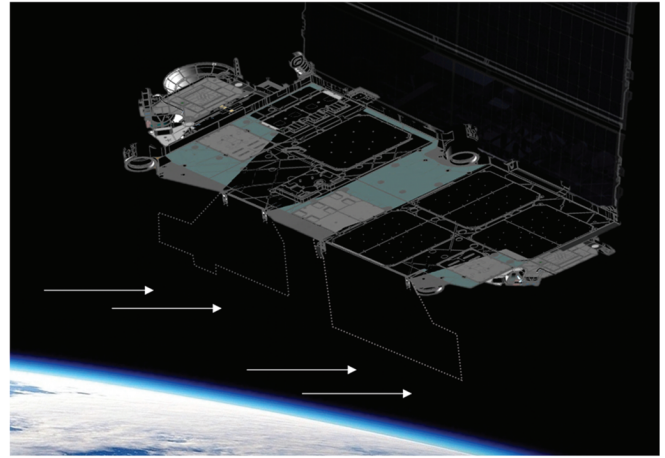

**Figure 7.** The Starlink v1.0 VisorSat satellites include shade visors which hang below and shade the nadir face of the satellite bus. Reproduced with permission from SpaceX 2020a. © SpaceX. Copyright 2023.

### 3.1.5. Starlink v1.5

The Starlink v1.5 satellites are a production version launched from 2021 June to 2022 December (Krebs 2022c). These satellites feature the addition of external laser modules for inter-satellite communication and do not include visors which would interfere with the laser communications. New dielectric mirror stickers cover the nadir-facing phased array antennae. The mirror stickers are intended to specularly reflect light into empty space and away from Earth. The parabolic dish antennae utilize the same dark material covers as the Starlink v1.0 VisorSat satellites. The solar array backing material is pigmented and darker than that of the previous Starlink satellites (SpaceX 2022).

## 3.2. OneWeb

The OneWeb satellites, shown in Figure 8, have a box-wing style design. The satellite bus is an approximately 1 m sized trapezoid shaped box wrapped in reflective multi-layer insulation. Two solar arrays extend on arms from opposite sides of the bus. The bus includes two angled antennae reflectors extended on arms from one side (Krebs 2022d).

The OneWeb satellites operate in a 1200 km circular orbit. The satellites initially launch into a lower orbit around 500 km and then individually use on-board thrusters to raise to the final operational orbit.

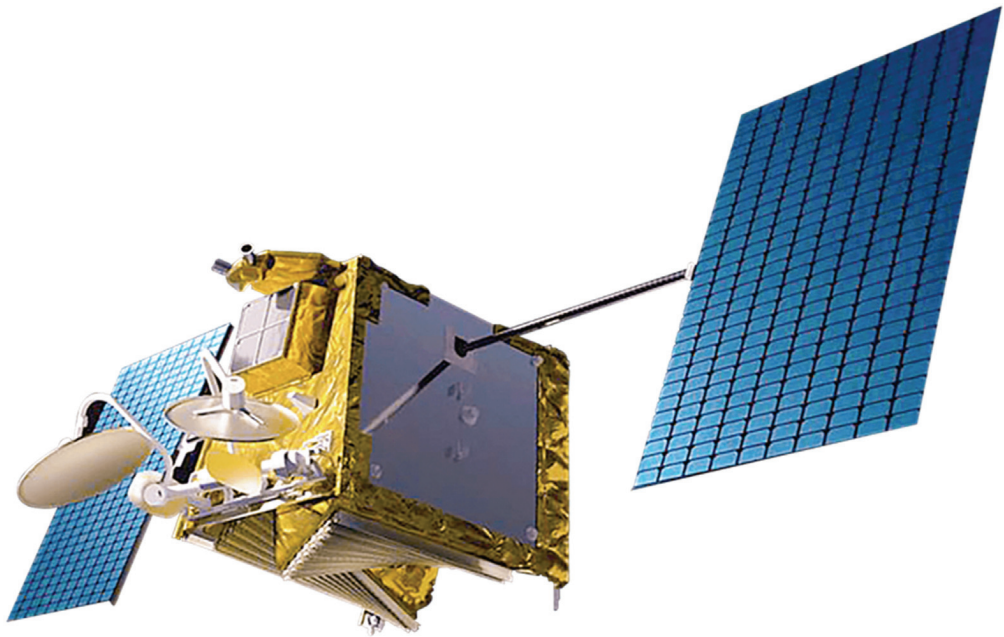

**Figure 8.** The OneWeb satellites utilize a box-wing style with a 1 m sized bus and two solar arrays. Reproduced with permission from Airbus OneWeb Satellites 2023. © Airbus OneWeb Satellites. Copyright 2023.

As of the end of 2022, there are 500 OneWeb satellites in orbit with an expected total of 648 satellites to complete the initial constellation. An extended constellation may include more than 6000 satellites. We do not know of any design or materials differences between OneWeb satellites and thus for analysis consider them all to be identical and a single population.

## 4. Results

From 2020 June to 2022 December the Steward Observatory LEO Satellite Photometric Survey produced over 16,000 photometric and astrometric measurements of nearly 2800 individual satellites from over 400 nights of observation. The breakdown of measurements is listed in Table 1. The photometric data presented here only includes measurements for which the satellite was at or near operational orbit (minimum 500 km for Starlink and 1000 km for OneWeb). Satellites that are not at operational orbit may be configured differently and may show different reflection characteristics and brightness. Additionally, we excluded measurements for which the position error is greater than 0.5° to avoid including potentially misidentified measurements.

The data is available for download from The University of Arizona ReData repository and use under the CC BY 4.0 license with appropriate citation of this publication (Krantz et al. 2023).

In addition to numerical statistics, we present the data visualized with sky-plots. The sky-plots show each individual measurement plotted as a dot at the position of observation on a projection of the sky looking up at zenith. The color of each dot signifies the value of the metric. The horizon is shown as a shaded boundary. The position of the Sun is plotted as a large yellow dot below the horizon.

The sky-plots show correlation of brightness with on-sky position and Sun-satellite-observer geometry. The Sun-satellite-observer geometry is defined by two vectors in 3D space with one vector pointing from the Sun to the satellite and one vector pointing from the satellite to the observer. Defining the pointing of these two vectors requires two angles, often described as the solar phase angle and the solar azimuth angle. Rather than plotting correlations with each angle separately, the sky-plots show correlation of brightness with both angles simultaneously without trying to uncouple the effects of one angle from the other, or deconvolving the effects of range or

**Table 1**
The Number of Observations for Each Satellite Population Included in the Photometric Analysis

| Satellite Population | Number of Satellites Observed | Number of Observations |
|---|---|---|
| Starlink v0.9 | 4 | 8 |
| Starlink v1.0 | 459 | 4724 |
| Starlink v1.0 DarkSat | 1 | 53 |
| Starlink v1.0 VisorSats | 924 | 4707 |
| Starlink v1.5 | 975 | 2539 |
| OneWeb | 356 | 4216 |

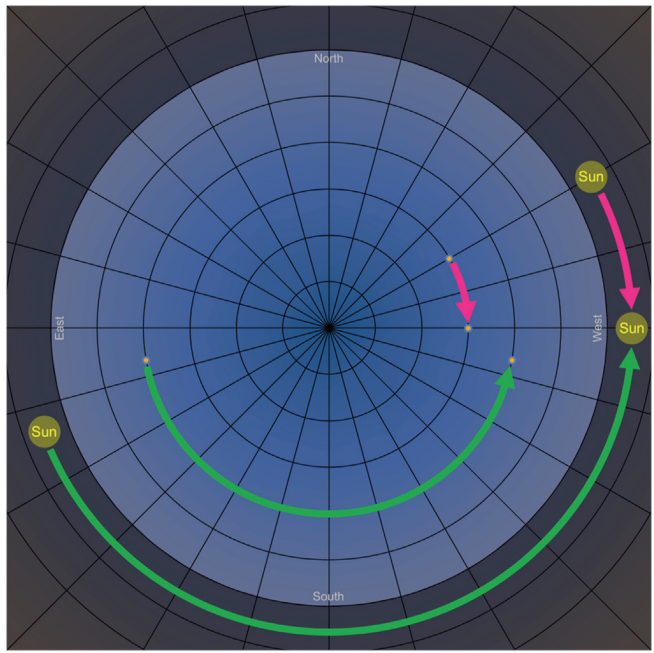

**Figure 9.** To standardize the Sun-satellite-observer geometry, the plotted position of each point is rotated around zenith such that the below-horizon Sun is at the same azimuth (West) for all measurements.

**Table 2**
The Visual Magnitude Statistics for Each Population of Satellites

| Satellite Population | N | Median | Brightest | Interdecile Range |
| --- | --- | --- | --- | --- |
| Starlink v0.9 | 8 | 5.06 | 2.42 | 2.42 → 8.35 |
| Starlink v1.0 | 4724 | 5.72 | 1.38 | 4.50 → 7.08 |
| Starlink v1.0 DarkSat | 53 | 6.52 | 4.03 | 4.52 → 7.70 |
| Starlink v1.0 VisorSat | 4707 | 6.87 | 1.56 | 5.25 → 8.15 |
| Starlink v1.5 | 2539 | 6.15 | 0.72 | 4.99 → 7.58 |
| OneWeb | 4216 | 8.23 | 2.86 | 7.05 → 9.10 |

**Note.** The interdecile range represents the typical range of brightness as seen in the sky.

airmass. In short, the sky-plots represent what an observer will see in the sky.

Throughout the year the Sun sets and rises at varying azimuths and thus the Sun-satellite-observer geometry is different for the same satellite observed at the same position on sky but in a different season. To accommodate this, we rotated the plotted position of each data point around zenith such that the Sun is at the same azimuth (West) for all measurements as demonstrated in Figure 9. This adjustment does not alter the geometry in a way that affects the apparent brightness of the satellite because the atmosphere is radially symmetric around the sky zenith point. It is possible that the satellite attitude is not consistent relative to the Sun or direction of orbit. Without specific knowledge of the satellite attitude, we must ignore this aspect and assume it is consistent.

Accommodating for the change in the Sun's below-horizon elevation in the same way is not possible because adjusting the plotted position in elevation would alter the geometry in a way that would change the apparent brightness. The change in elevation would correspond to a change in satellite range and atmospheric airmass and thus a change in apparent brightness. During a typical observation period the Sun moves up to 20° in elevation. To investigate if this change in elevation has a significant effect on the pattern of apparent brightness we produced sky-plots with data from consecutive limited ranges of Sun elevations (i.e., −5° to −10°, −10° to −15° etc.). We did not see clear or significant differences in the pattern of apparent brightness and thus decided to ignore the changes in the Sun's elevation though this may warrant further examination in the future.

The sky-plots are foremost a visualization of what we observed from the Tucson, AZ area (∼32°N). We do not intend them to be a universal representation for all locations though we expect the apparent brightness and reflection patterns will be very similar for observers across the globe. Observers at high latitudes may see more significant differences due to the different distribution of visible satellites in the sky.

### 4.1. Visual Magnitude

The visual magnitude statistics for each satellite population are listed in Table 2 and histograms shown in Figure 10. The distribution of values is not statistical randomness but the range of apparent brightness that each population exhibits. This range is quantified numerically by the interdecile range which signifies the typical range of brightness for the satellites as seen in the sky.

Comparing the histograms and interdecile ranges immediately shows differences between the satellite populations. The OneWeb satellites are notably fainter than all the Starlink satellites, largely due to their higher orbit and thus farther range as seen in the sky. The different Starlink populations show minor differences in typical brightness with the Starlink v1.0 VisorSat satellites appearing approximately 1 mag fainter than the Starlink v1.0 satellites. The Starlink v1.5 satellites are slightly brighter than the Starlink v1.0 VisorSat satellites but not as bright as the Starlink v1.0 satellites, indicating that the change of material finishes did reduce the apparent brightness but not as much as the visors did.

While the interdecile range signifies the typical range of brightness each satellite population exhibits as seen in the sky, all populations included instances where the satellites appeared significantly brighter. These are all instances where the satellites exhibited flares due to specular reflections of sunlight. While not descriptive of their typical behavior it demonstrates satellites notably reduce apparent brightness on the half of the sky near the below-horizon Sun but did not reduce the apparent brightness on half of the sky opposite the Sun. The OneWeb satellites do not show as much distinctive pattern but are also brightest just below zenith opposite the Sun and show a similar hot spot at low solar elongation above the below-horizon Sun.

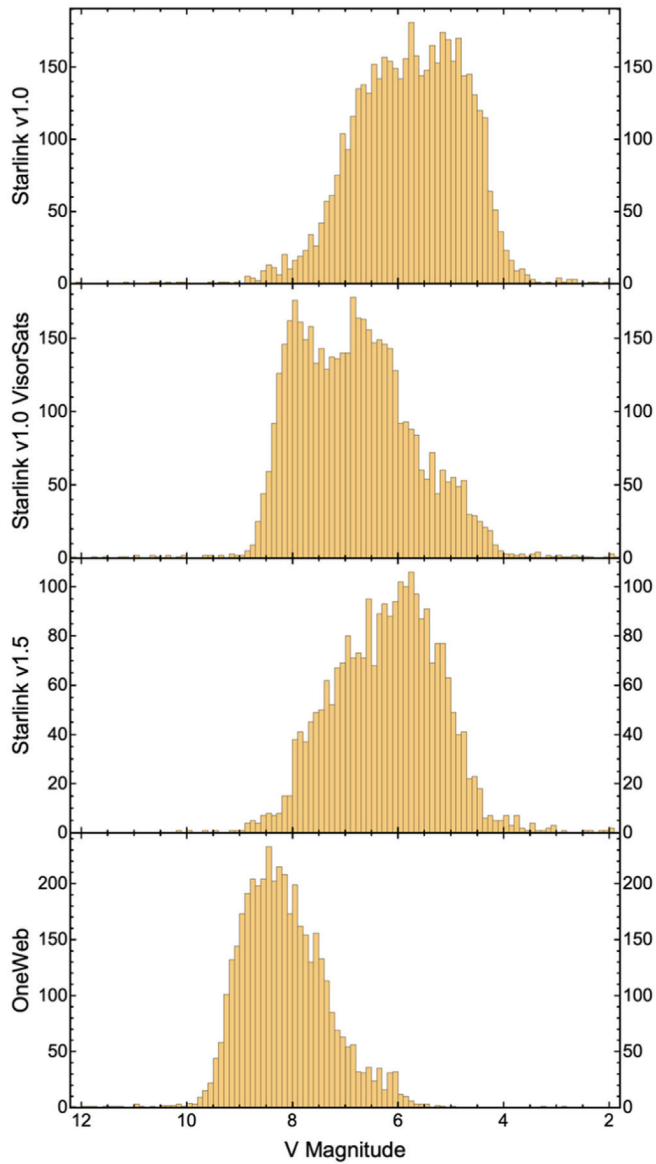

**Figure 10.** Histograms show the typical range of visual magnitudes for each satellite population as seen in the sky. The distributions are not statistical randomness but the range of apparent brightness that each population exhibits.

### 4.2. Expected Photon Flux

For evaluating the brightness of moving satellites in the context of impact on astronomy, visual magnitudes are an insufficient metric. The magnitude metric describes an instantaneous brightness irrespective of observing tools or methodology. For example, consider an image with an 8th magnitude star. A second image with a longer exposure will show the same star with a higher flux count, but it will still measure as 8th magnitude. The same is not true for a moving satellite which appears as a trail or line through the image. While a longer exposure time will increase the flux count for stars, the satellite trail will always appear the same with the same flux count regardless of image exposure time. The flux count for the satellite trail depends on how bright it is and how long it takes to cross one pixel on the imaging sensor. This effective exposure time is dependent on the plate scale of the sensor, i.e., the angular size of each pixel, and the satellite's apparent on-sky velocity, which in turn is dependent on the satellite's range and orbit. A slow-moving faint satellite may appear just as bright, per pixel, as a faster moving brighter satellite.

To evaluate the impact on astronomy we want to quantify satellite brightness as seen by an image sensor. To do this we calculated the expected photon flux per arcsec of satellite trail as seen by a theoretical telescope and image sensor with 1 m² of open aperture, 1″ pixels, 100% quantum efficiency, a gain a 1 ADU e$^{-1}$, and no systemic losses. The calculated numbers loosely represent the photon flux per pixel for a 1 m class telescope, assuming the satellite trail width is 1″. Following Equation (1), we converted each photometric measurement from magnitude units to photon flux using a reference zero magnitude flux, $F_0$, of $8.46175 \times 10^9$ photons m$^{-2}$ s$^{-1}$ (Bessell et al. 1998) and then scaled each measurement to the effective exposure time for the satellite to move 1″.

$$\text{Photon Count} = 10^{-0.4 m_v} * F_0 * \frac{1''}{\text{Satellite Angular Velocity}} \quad (1)$$

Our intention is to provide an intuitive view of whether or not a satellite will be bright enough to saturate a sensor (∼65k counts for a 16-bit sensor) or otherwise significantly impact an instrument or project. Even if the satellite trail does not saturate the sensor, the Poisson photon noise may be large enough to completely obscure an overlapped faint science target. To assess the potential impact on telescopes and instruments scale the numbers up or down according to the sensor specifications.

that every satellite type has the potential to be very bright if seen in the geometry to produce a specular flare.

The sky-plots in Figure 11 show the correlation of apparent brightness with on-sky position and Sun-satellite-observer geometry for each population of satellites. They show how the typical range of brightness distributes on the sky and where the satellites appear brightest, which is notably not at zenith where they are closest in range. The Starlink satellites all appear brightest at mid-elevations opposite the Sun with an additional hot spot at low solar elongation above the below-horizon Sun. The visors included on the Starlink v1.0 VisorSat

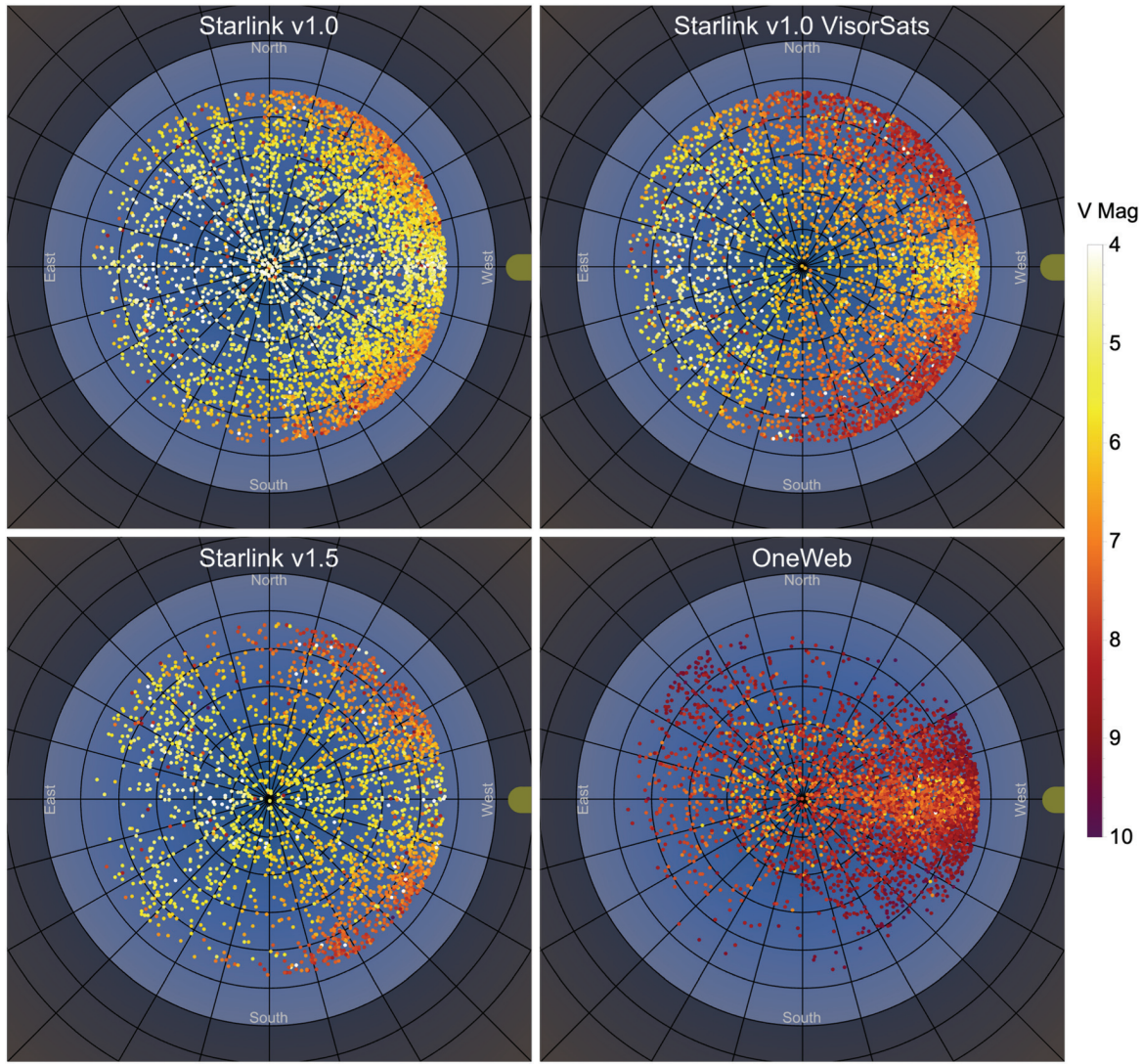

**Figure 11.** The visual magnitude sky-plots for each satellite population show the correlation of apparent brightness with on-sky position relative to the below-horizon Sun. To standardize the Sun-satellite-observer geometry, the plotted position of each point is rotated around zenith such that the below-horizon Sun is at the same azimuth (West) for all measurements.

For example, for a 4 m telescope, multiply the numbers by a factor of ∼12 due to the larger aperture. For an instrument with 3″ pixels, multiply the numbers by a factor of 3 due to the longer time a satellite takes to cross each pixel, or vice versa for smaller pixels. Similarly, if the satellite trail width is larger than 1″ due to seeing conditions or defocused appearance, divide the numbers by the apparent trail width. For a sensor with gain of 4 ADU e$^{-1}$, divide the numbers by 4 to account for the lower conversion ratio.

Anyone wanting to evaluate the impact satellites will have on their instrument or observing project should consider the expected photon flux measurements instead of visual magnitudes, as the expected photon flux measurements consider the satellite's apparent on-sky velocity and real brightness as seen by the sensor.

The expected photon flux statistics for each satellite population are listed in Table 3 and histograms shown in Figure 12. All populations of Starlink satellites cover a broad range of expected photon flux with the majority of satellites not appearing bright enough to cause severe impacts. However, the brighter end of the typical range does pose a threat. The Starlink v1.0 satellites' interdecile range top end is 67k counts, enough to saturate the sensor in a 1 m class telescope. The Starlink v1.0 VisorSat and Starlink v1.5 satellites are fainter with around 40k counts at the top end of the interdecile range, not enough to saturate the sensor in a 1 m class telescope, but

**Table 3**
The Expected Photon Flux Statistics for Each Satellite Population

| Satellite Population | N | Median | Brightest | Interdecile Range |
| --- | --- | --- | --- | --- |
| Starlink v0.9 | 8 | 55 k | 629 k | 2.7 k → 629 k |
| Starlink v1.0 | 4724 | 28 k | 1062 k | 9.6 k → 67 k |
| Starlink v1.0 DarkSat | 53 | 11 k | 158 k | 4.3 k → 66 k |
| Starlink v1.0 VisorSat | 4707 | 8.8 k | 1166 k | 3.5 k → 41 k |
| Starlink v1.5 | 2539 | 15 k | 3250 k | 5.6 k → 45 k |
| OneWeb | 4216 | 4.7 k | 661 k | 2.5 k → 14 k |

**Note.** These numbers loosely represent how bright a 1″ wide satellite trail will appear in a recorded image for a theoretical 1 m class telescope.

enough to likely cause significant detriment to the data. The OneWeb satellites are fainter than all the Starlink satellites with around 14k counts at the high end of the interdecile range. This is likely below the threshold for severe impacts with a 1 m class telescope, though not for larger telescopes. All satellite populations showed the potential for severe impacts with the brightest satellites observed exhibiting expected photon flux which far exceed the saturation threshold for a 1 m class telescope.

The sky-plots in Figure 13 show the correlation of expected photon flux with on-sky position and Sun-satellite-observer geometry for each population of satellites. While similar in appearance to Figure 11, there are important differences. Satellites around zenith now appear fainter because satellites at zenith move faster on-sky, have a shorter effective exposure time, and thus appear less bright to an imaging sensor. Conversely, satellites lower in the sky move slower and appear brighter.

The Starlink v1.0 satellites threaten to severely affect sensors in a 1 m class telescope with expected photon flux greater than 30k across nearly the entire sky. With expected photon fluxes around 10k or less, the Starlink v1.0 VisorSat and Starlink v1.5 satellites do not pose a severe threat on the half of the sky near the below-horizon Sun, but do on the half of the sky opposite the Sun being almost as bright as the Starlink v1.0 satellites. However, it is important to consider that satellites in the half of the sky opposite the Sun will be the first to be eclipsed by the Earth as the Sun moves further below the horizon and twilight ends.

In comparison, the OneWeb satellites do not pose a severe threat across most of the sky except in small areas where they may appear brighter. The OneWeb satellites are in a higher orbit than the Starlink satellites and thus have a slower angular on-sky velocity and a longer effective exposure time. Despite this, the OneWeb satellites appear generally less bright in terms of expected photon flux than the Starlink satellites.

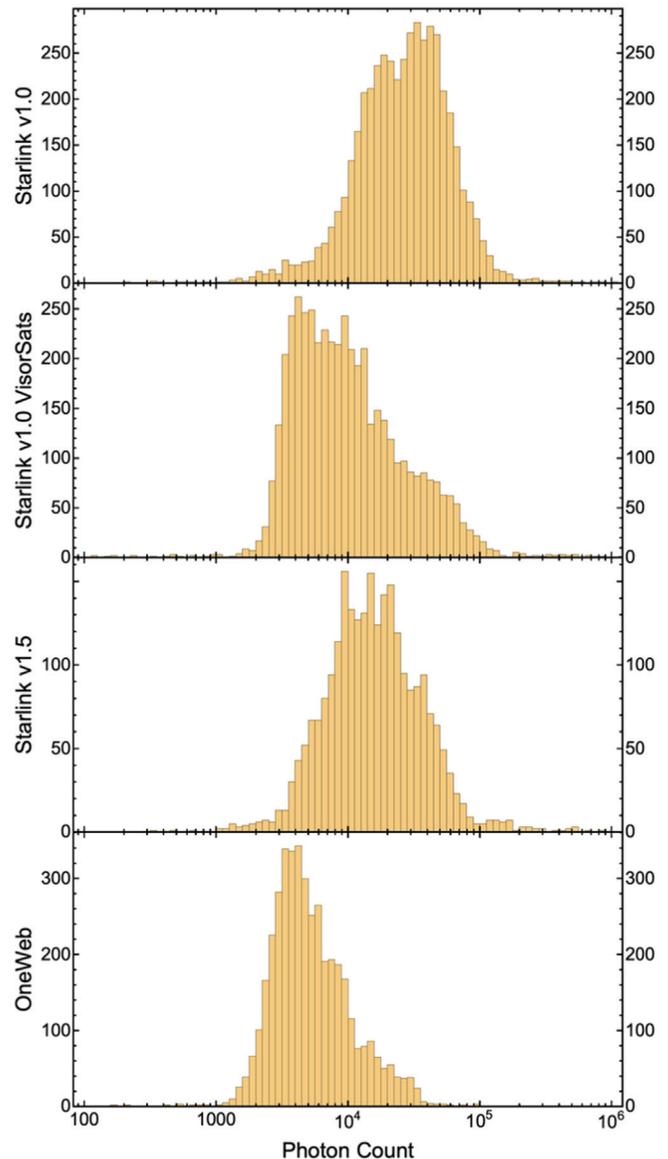

**Figure 12.** Histograms show the typical range of expected photon flux for each satellite population. These numbers loosely represent how bright a 1″ wide satellite trail will appear in a recorded image for a theoretical 1 m class telescope.

### 4.3. Effective Albedo

Visual magnitude and the expected photon flux metrics describe the apparent brightness of the satellites but do not serve well for evaluating the reflection properties of these satellites. The apparent brightness metrics describe how bright the satellites appear to an observer but do not describe how reflective the satellites are or how the apparent reflectivity changes with viewing geometry. We need a different metric to quantify where a satellite is reflecting more or less light and evaluate how this correlates with geometry and satellite design.

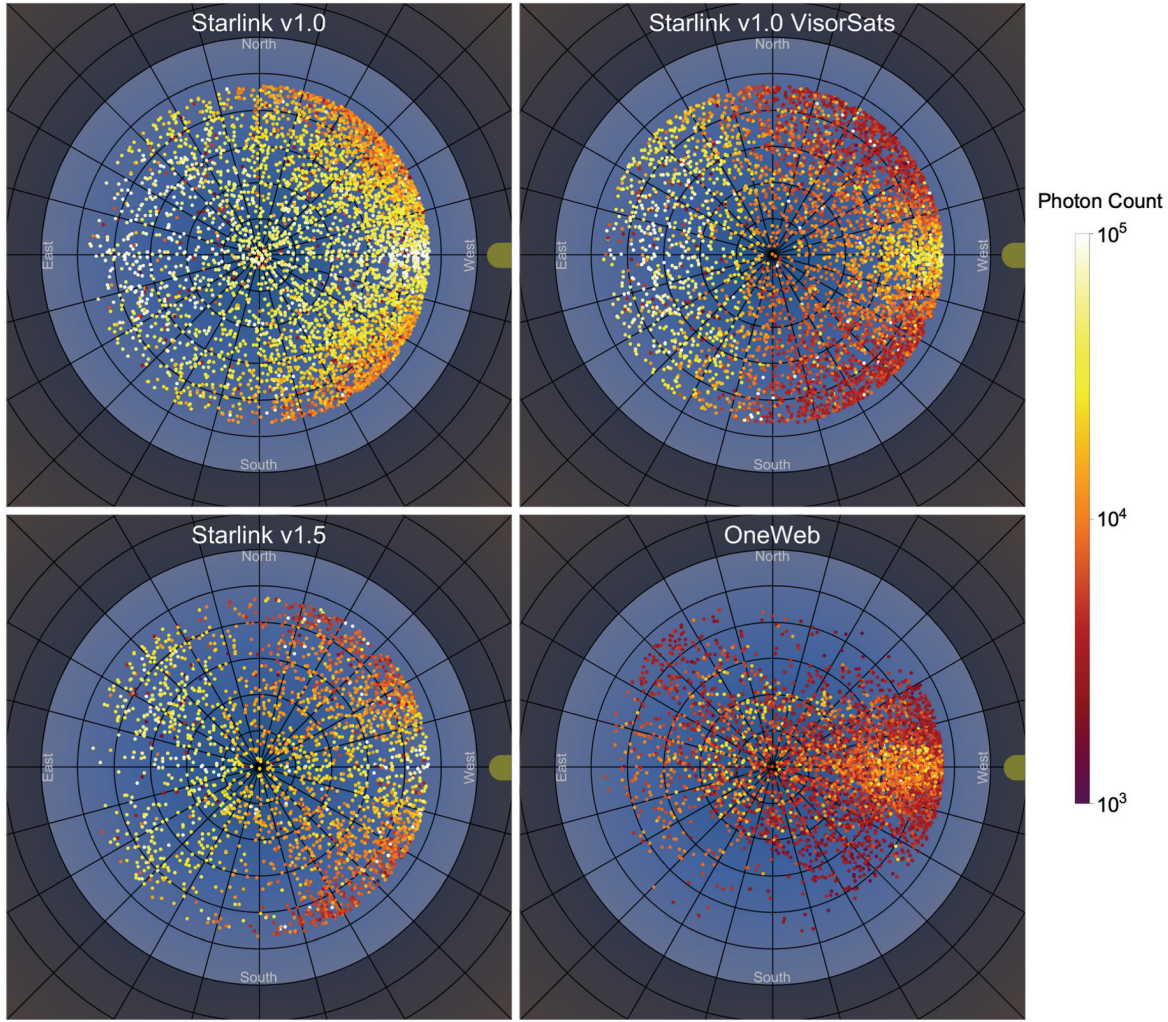

**Figure 13.** The sky-plots for each satellite population show the correlation of expected photon flux with on-sky position relative to the below-horizon Sun. These numbers loosely represent how bright a 1″ wide satellite trail will appear in a recorded image for a theoretical 1 m class telescope. In comparison to Figure 11, satellites around zenith now appear fainter because satellites at zenith move faster on-sky, have a shorter effective exposure time, and thus appear less bright to an imaging sensor. Conversely, satellites lower in the sky move slower and appear brighter. The expected photon flux is more relevant to assessing the impact to research astronomy than visual magnitude measurements. To standardize the Sun-satellite-observer geometry, the plotted position of each point is rotated around zenith such that the below-horizon Sun is at the same azimuth (West) for all measurements.

One approach is to fit the data to a bidirectional reflectance distribution function (BRDF). However, doing so in an informative way is difficult without detailed knowledge of the satellite design, shape, size, and materials. A BRDF is only applicable to a specific satellite design and is not universally applicable. In the interest of simple and universal analysis, we instead focus on a parameter which broadly describes the reflecting properties of a space object: its albedo.

The simplest and most common method for estimating the apparent brightness of a reflecting space object is the diffuse sphere model. This model treats the space object as a gray uniform sphere with perfect Lambertian reflectance. Then accounting for the object's range $R$, phase angle $\theta$, size $A$, and albedo $\rho$ calculates the expected apparent brightness $m_v$ in Equation (2) (Hejduk 2011).

$$m_v = m_{\text{Sun}} - 2.5 \log\left(\frac{2}{3\pi^2} A \rho [(\pi - \theta)\cos\theta + \sin\theta]\right) + 5 \log(R) \quad (2)$$

The diffuse sphere model is adequate for estimating the brightness of space objects but does not accurately predict the brightness of many satellites in many geometries because satellites are not uniform diffuse spheres. Satellites are complex objects with non-spherical structure and many discrete surfaces. The combination of multiple reflections from surfaces

**Table 4**
The Effective Albedo Statistics for Each Satellite Population

| Satellite Population | N | Median | Max | Interdecile Range |
| --- | --- | --- | --- | --- |
| Starlink v0.9 | 8 | 1.79 | 132.22 | 0.11 → 132.22 |
| Starlink v1.0 | 4724 | 1.47 | 251.91 | 0.61 → 6.24 |
| Starlink v1.0 DarkSat | 53 | 0.62 | 12.23 | 0.20 → 1.90 |
| Starlink v1.0 VisorSat | 4707 | 0.53 | 56.99 | 0.17 → 2.87 |
| Starlink v1.5 | 2539 | 0.70 | 240.97 | 0.30 → 3.69 |
| OneWeb | 4216 | 0.55 | 86.56 | 0.15 → 1.86 |

**Note.** The effective albedo is a relative measure of satellite reflectivity in comparison to an idealized 1 m sphere in the same illumination geometry.

with both diffuse and specular properties produces the sum of reflected light that we observe as the apparent brightness. Even slight changes in illumination geometry or satellite orientation can yield large changes in apparent brightness, which the diffuse sphere model does not include.

Here we utilize the diffuse sphere model in a different way. We use it as a known benchmark. We know how an ideal uniform diffuse sphere will reflect light and compare the observed satellite to it. After inverting Equation (2) to create Equation (3), we input our measured apparent brightness $m_v$, range $R$, phase angle $\theta$, assign a default unit size for $A$, and calculate the albedo. This produces what we call the *effective albedo*, $\rho_{\text{eff}}$. It is the albedo needed for an ideal diffuse sphere of unit size to produce the observed brightness in the observed geometry.

$$\rho_{\text{eff}} = \frac{3\pi^2 R^2 10^{\frac{m_{\text{Sun}} - m_v}{2.5}}}{2[(\pi - \theta)\cos\theta + \sin\theta]} \quad (3)$$

The effective albedo is not a real measurement of albedo and should not be interpreted as such. The utility in this metric is the relative change in value across different reflecting geometries. A higher than baseline effective albedo indicates an increase in reflectivity which is directly inherent to the shape, design, and materials of the satellite. It shows where a satellite is brighter than it could be assuming that a diffuse reflection is the minimum brightness.

The effective albedo statistics for each satellite population are listed in Table 4 and histograms shown in Figure 14. The design modifications implemented by SpaceX with the Starlink v1.0 VisorSat and Starlink v1.5 satellites reduced the typical effective albedo by more than 50% compared to the Starlink v1.0 satellites. The typical effective albedo for the OneWeb satellites is similar to that of the Starlink v1.0 VisorSat satellites indicating that if the two populations were at the same orbit height, they would likely appear similar in brightness.

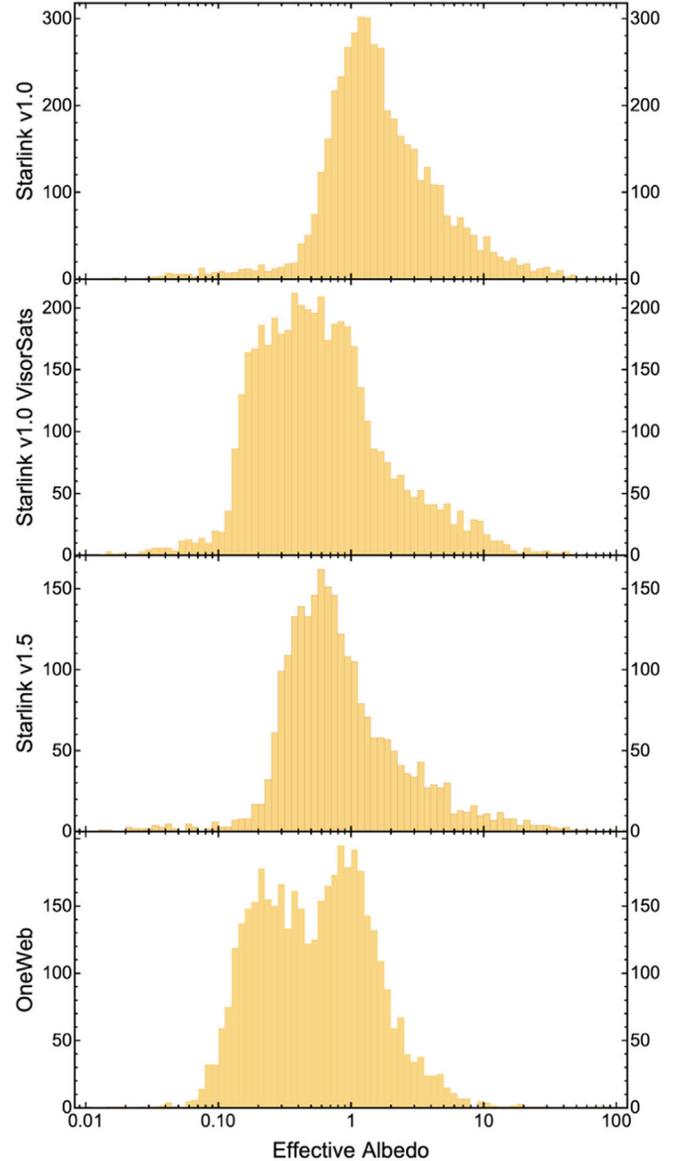

**Figure 14.** Histograms show the typical range of effective albedo for each satellite population. The effective albedo is a relative measure of satellite reflectivity in comparison to an idealized 1 m sphere in the same illumination geometry.

The sky-plots in Figure 15 show the correlation of effective albedo with on-sky position and Sun-satellite-observer geometry for each population of satellites. All the satellite populations exhibit the highest effective albedo at low solar elongations above the below-horizon Sun. This is indicative of specular low-angle glancing reflections off the satellite's nadir faces. This effect is particularly prominent with the Starlink satellites which all have a large flat nadir face. The effects of the visors on the Starlink v1.0 VisorSat satellites are clearly seen as a broad darker region across the sky perpendicular to the direction of sunlight. The Starlink v1.5 satellites show a similar darkening, possibly a result of the mirror stickers reflecting light away from Earth at moderate elongation angles. All the Starlink satellites show increased effective albedo opposite the Sun indicating back-reflection off the solar array. The darker pigment used in the solar array backing with the Starlink v1.5 satellites appears to have reduced the back-reflection by a small amount.

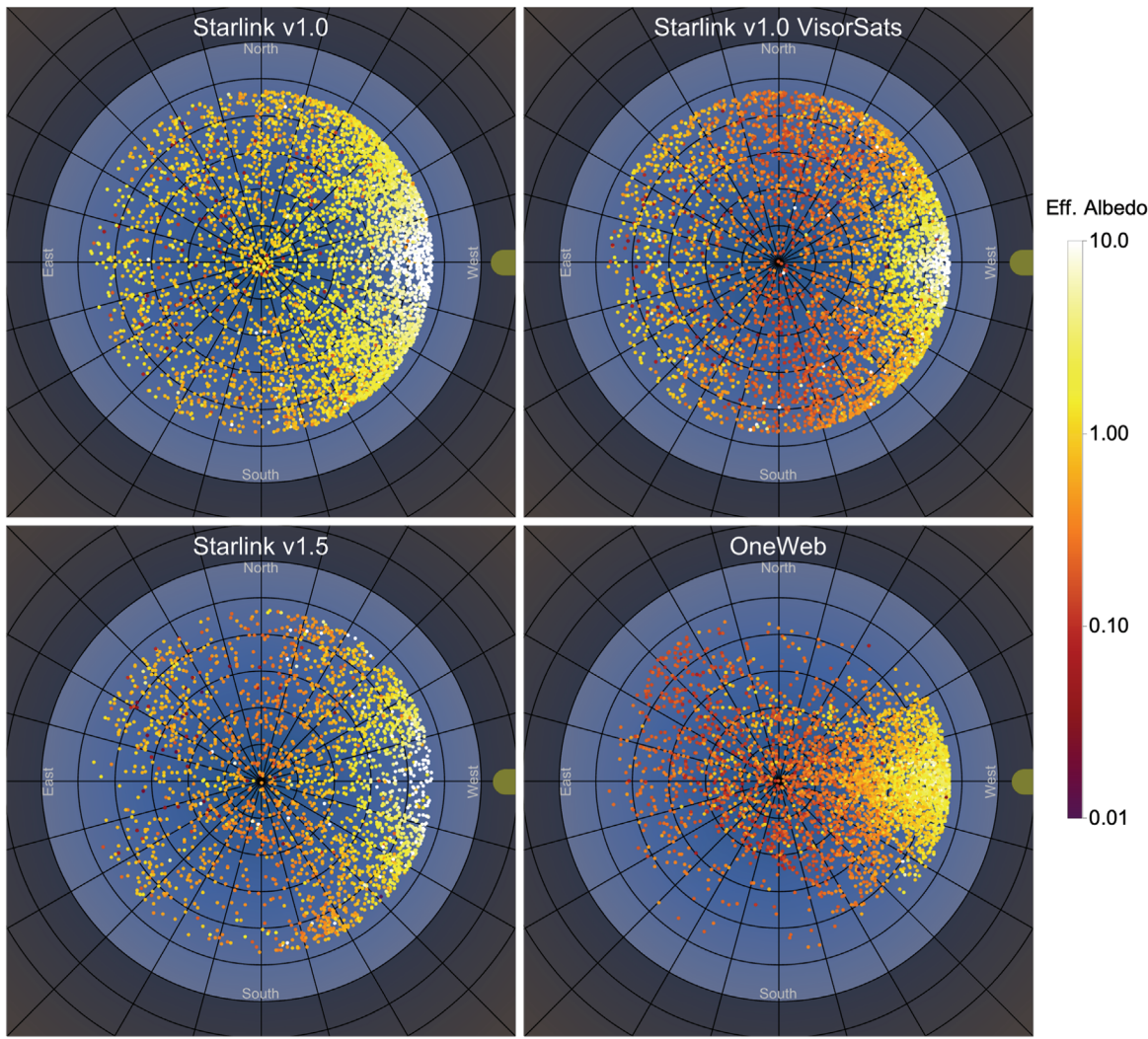

**Figure 15.** The sky-plots for each satellite population show the correlation of effective albedo with on-sky position relative to the below-horizon Sun. The effective albedo is a relative measure of satellite reflectivity in comparison to an idealized 1 m sphere in the same illumination geometry. To standardize the Sun-satellite-observer geometry, the plotted position of each point is rotated around zenith such that the below-horizon Sun is at the same azimuth (West) for all measurements.

Besides the hot spot at low solar elongation, the OneWeb satellites do not show any other distinct areas of higher effective albedo indicating the satellites apparent brightness is largely driven by diffuse or diffuse-like reflections. The wrinkles in reflective foil on the OneWeb satellite bus possibly scatter light in many directions which appear diffuse-like.

### 4.4. Astrometric Position and TLE Accuracy

During image processing we determine the observed astrometric position of each satellite at the time of observation. We also compute the expected position of the satellite from the most recent TLE prior to the time of observation. We measure the difference between the real position and expected position to evaluate the TLE accuracy. We do this comparison for both the standard issue TLEs produced by the US Space Force and the Supplemental TLEs produced by Celestrak.com from first party telemetry data provided by the satellite operators (Celestrak 2022).

We only performed a simple analysis of TLE accuracy and did not differentiate between along-track and cross-track position differences. Our intention is to provide a general assessment of

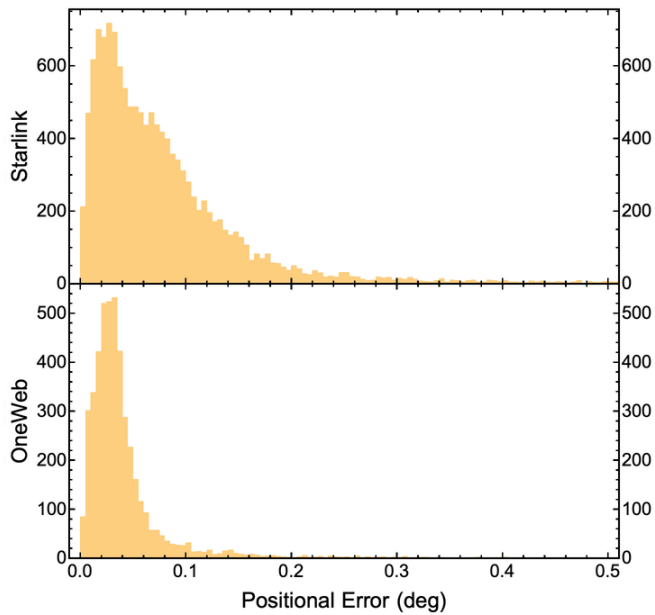

**Figure 16.** The Supplemental TLE positional error histograms determined by comparing the observed satellite position to that calculated from the Supplemental TLEs.

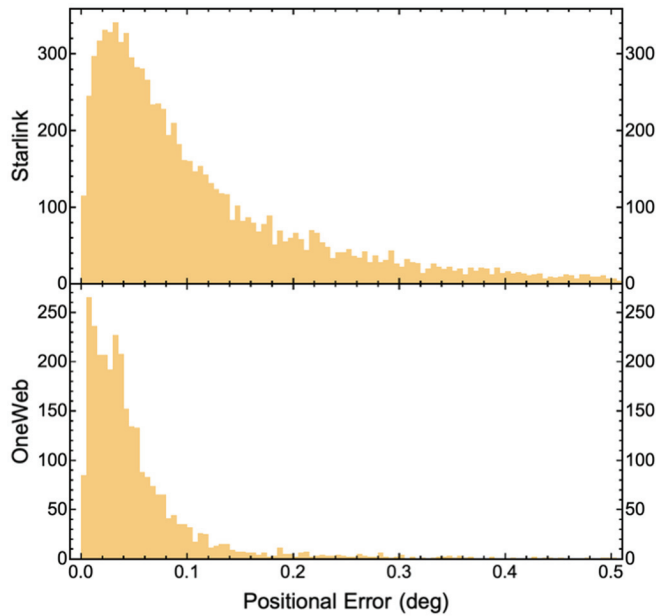

**Figure 17.** The Standard Issue TLE positional error histograms determined by comparing the observed satellite position to that calculated from the Standard Issue TLEs.

**Table 5**
The TLE Accuracy Statistics for Each Satellite Population and TLE Source

| Satellite Population | TLE Source | N | Median | Interdecile Range |
|---|---|---|---|---|
| Starlink | Celestrak Supplemental | 13442 | 0.06° | 0.02° → 0.17° |
|  | Standard Issue | 9034 | 0.08° | 0.02° → 0.31° |
| OneWeb | Celestrak Supplemental | 4599 | 0.03° | 0.01° → 0.07° |
|  | Standard Issue | 2978 | 0.04° | 0.01° → 0.12° |

**Note.** We measured the angular separation between the observed satellite position and the expected position calculated from the TLE.

satellite position prediction accuracy that is representative of what a general-purpose observatory will see without highly critical consideration of clock accuracy and instrument timing. We have not performed a critical assessment of the clock or shutter timestamping of our observatory which utilizes an onsite time server. Small timing errors of only 50 ms could produce a 1.5′ position difference with a satellite moving 0.5 deg per second on sky. An in-depth and accurate study of satellite position and prediction accuracy will be a significant work requiring a foundation of robust systematic timing accuracy.

The statistics for TLE accuracy for each satellite population and TLE source are listed in Table 5 and histograms shown in Figures 16 and 17. Overall, TLE accuracy is very good with most satellites appearing within a few arcminutes of the expected position. The Supplemental TLEs are frequently more accurate with a tighter distribution and shorter tail in the histogram. However, the Supplemental TLEs are not absolutely more accurate. The best Supplemental TLE results and Standard Issue TLE results show the same level of accuracy. The Supplemental TLEs are published more frequently than the Standard Issue TLEs so the better accuracy may be due to the availability of a more contemporaneous TLE. The Supplemental TLEs also account for planned maneuvers and include future trajectories planned by the operator. This is in contrast to the Standard Issue TLEs which are created from observational data and thus always fit to the trajectory at the time of observation.

For both TLE sources, the OneWeb satellites exhibit better accuracy than the Starlink satellites. This is likely due to their higher orbit and farther observed range. Uncertainty in absolute position calculated by the SGP4 model, typically around 1 km, is diminished with increased range when transformed to a topocentric pointing. Additionally at the higher orbit there is less drag and less variation in drag, thus the SGP4 model predictions are more reliable.

## 5. Conclusions

The Steward Observatory LEO Satellite Photometric Survey has operated for over two years and in that time produced over 16,000 photometric and astrometric measurements of nearly 2800 Starlink and OneWeb satellites. With an all-sky, all-geometries approach, we captured the full range of apparent brightness for each population of satellites and correlated the distribution of brightness with on-sky position and relative Sun-satellite-observer geometry. Each population of satellites exhibits a distinct pattern of brightness and notably do not appear brightest near zenith.

A new metric, the expected photon flux, assesses the potential impact on astronomy by quantifying the expected flux count for a satellite trail as seen by a sensor in a theoretical 1 m class telescope. This metric provides an intuitive view to determine if the satellites are bright enough to saturate a detector or cause severe impact on the data. All Starlink and OneWeb satellites will easily be seen by a 1 m class telescope. The Starlink satellites at the top end of the typical brightness range threaten to saturate or severely affect sensors in a 1 m class telescope.

A second new metric, the effective albedo, reveals geometries where the satellites are more reflective than baseline. This ties to the physical structure of the satellite and enables one to infer which elements of the satellite are responsible for the increased reflectivity. The various modifications SpaceX made to the Starlink satellite design reduced effective albedo in certain geometries but not in all.

Overall, the changes implemented by SpaceX reduced the quantity of reflected light by more than 50% as measured with the effective albedo. However, this only reduced the apparent brightness by 0.5–1.0 visual magnitudes. Both the Starlink v1.0 VisorSat and Starlink v1.5 satellites typically appear as bright as 5th magnitude. Reducing the apparent brightness by an additional 2 mag, to meet the 7th magnitude threshold called for by the IAU CPS, corresponds to reducing the quantity of reflected light by an additional 84%, a significant engineering challenge for spacecraft design.

Current mitigation efforts primarily focus on reducing the apparent brightness of satellites by changing their design, configuration, and orbit. However, changing the satellites is not the only option to mitigate the impacts. New software tools and image processing methods could remove or mask satellite trails and recover images that would otherwise be discarded. Strategically scheduling observations could avoid satellites in the sky, or at least avoid the brightest satellites and regions of the sky with the most visible satellites. Additional hardware, such as fast shutters, could block light from satellites reaching the detector. New instrumentation design needs to accommodate the appearance of bright satellites and strategically work around them or at least operate without severe side effects. New observing programs need to consider the impacts and risks posed by satellite interference when planning methodology and submitting proposals.

As recognized at the SATCON1 workshop, no single solution will entirely mitigate all the impacts, nor is any single group responsible for producing all the solutions. The best solution will be a combination of mitigation efforts from multiple domains. The IAU CPS is working to bring astronomers, satellite operators, and other communities together to establish reasonable guidelines and support the development of mitigation efforts.

The results of the Steward Observatory LEO Satellite Photometric Survey provide a complete view of current satellite brightness as seen in the sky. This baseline is a reference from which future brightness reducing mitigation efforts can be compared. Observers can use the sky-plots of apparent brightness and expected photon flux metric to evaluate the potential impact on their projects and possibly strategically execute observations to avoid the brightest satellites. Satellite operators can use the effective albedo sky-plots to understand the geometries where their satellites are reflecting more light than baseline and use this information to evaluate their satellite design and identify opportunities to modify the design to reduce apparent brightness.

### Acknowledgments

We thank the staff of the Steward Observatory Mountain Operations division for their valuable ongoing support at the Mt. Lemmon Sky Center. We also thank the staff of Biosphere 2 for their support during our temporary relocation.

### ORCID iDs

Harrison Krantz  https://orcid.org/0000-0003-0000-0126